\documentclass[aip,graphicx]{revtex4-1}

\draft 

\usepackage{graphicx}
\usepackage{dcolumn}
\usepackage{bm}

\usepackage[utf8]{inputenc}
\usepackage[T1]{fontenc}
\usepackage{mathptmx}
\usepackage{etoolbox}
\usepackage[normalem]{ulem}
\usepackage{xcolor}

\makeatletter
\def\@email#1#2{%
 \endgroup
 \patchcmd{\titleblock@produce}
  {\frontmatter@RRAPformat}
  {\frontmatter@RRAPformat{\produce@RRAP{*#1\href{mailto:#2}{#2}}}\frontmatter@RRAPformat}
  {}{}
}%
\makeatother

\begin{document}


\title{Influence of the downstream blade sweep on cross-flow turbine performance}
\author{A. Snortland}
 \email{abigales@uw.edu}
 
\author{A. Hunt}%
\affiliation{ 
University of Washington Department of Mechanical Engineering 3900 E Stevens Way NE, Seattle, WA 98195, United States
}%

\author{O. Williams}
\affiliation{
University of Washington Department of Aeronautics and Astronautics 3940 Benton Lane NE, Seattle, WA 98195, United States}%

\author{B. Polagye}%
\affiliation{ 
University of Washington Department of Mechanical Engineering 3900 E Stevens Way NE, Seattle, WA 98195, United States
}%

\date{\today}

\textbf{This article may be downloaded for personal use only. Any other use requires prior permission of the author and AIP Publishing. This article appeared in (A. Snortland, A. Hunt, O. Williams, B. Polagye; Influence of the downstream blade sweep on cross-flow turbine performance. J. Renewable Sustainable Energy 1 January 2025; 17 (1): 013301.) and may be found at (https://doi.org/10.1063/5.0230563).}

\begin{abstract}
Cross-flow turbine (known as vertical-axis wind turbines or ``VAWTs'' in wind) blades encounter a relatively undisturbed inflow for the first half of each rotational cycle (``upstream sweep'') and then pass through their own wake for the latter half (``downstream sweep''). While most research on cross-flow turbine optimization focuses on the power-generating upstream sweep, we use single-bladed turbine experiments to show that the downstream sweep strongly affects time-averaged performance. We find that power generation from the upstream sweep continues to increase beyond the optimal tip-speed ratio. In contrast, the downstream sweep consumes power beyond the optimal tip-speed ratio due to unfavorable lift and drag directions relative to rotation and a potentially detrimental pitching moment arising from rotation-induced virtual camber. Downstream power degradation increases faster than upstream power generation, such that downstream sweep performance determines the optimal tip-speed ratio. In addition to performance measurements, particle image velocimetry data is obtained inside the turbine swept area at three tip-speed ratios. This illuminates the mechanisms underpinning the observed performance degradation in the downstream sweep and motivates an analytical model for a limiting case with high induction. Performance results are shown to be consistent across 55 unique combinations of chord-to-radius ratio, preset pitch angle, and Reynolds number, underscoring the general significance of the downstream sweep.
\end{abstract}

\pacs{}

\maketitle 

\section{Introduction}


Cross-flow turbine technologies are able to harness the kinetic energy of wind, tidal currents, and rivers. In comparison to axial-flow turbines, cross-flow turbines (``vertical-axis'' in wind), operate at lower rotation rates (reduced noise, lower risk of animal collision), are insensitive to inflow direction (obviating yaw control), have lower blade bending stresses and simpler construction, support favorable generator positioning (e.g., on the surface with floating platforms) \cite{VAWTreview,WindComp}. For these reasons, cross-flow turbines may have advantages over axial-flow turbines across a wide range of scales and applications. However, because cross-flow turbines rotate perpendicular to the inflow, the blades experience more complex fluid dynamics that vary substantially throughout each rotation. Consequently, the drivers of cross-flow turbine performance are less well understood, especially during the second half of the rotation where the inflow is substantially modified by power extraction in the first half of the rotation. The first half of the blade rotation (the ``upstream sweep'') is commonly referred to as the ``power stroke'' as it accounts for the majority of power generation, while the second half of the rotation (the ``downstream sweep'') is characterized by limited power production, post- and secondary-stall events, and boundary layer reattachment \cite{Snortland2023,sebstalldilema}. 
Prior works \cite{Me,REZAEIHA2018TSRimpact,Snortland2023,Mukul,sebstalldilema} have found substantial differences in performance between the upstream and downstream sweeps. The upstream sweep clearly influences overall turbine efficiency, but our knowledge of cross-flow turbine operation is fundamentally incomplete without a full understanding of the downstream sweep.  
Here we specifically investigate the downstream sweep, which has received significantly less dedicated research attention. In doing so, we aim to illuminate its importance on cross-flow turbine performance and to clarify the drivers of downstream sweep performance.

\begin{figure}[t!]
\centering
    \includegraphics[width=0.9\linewidth]{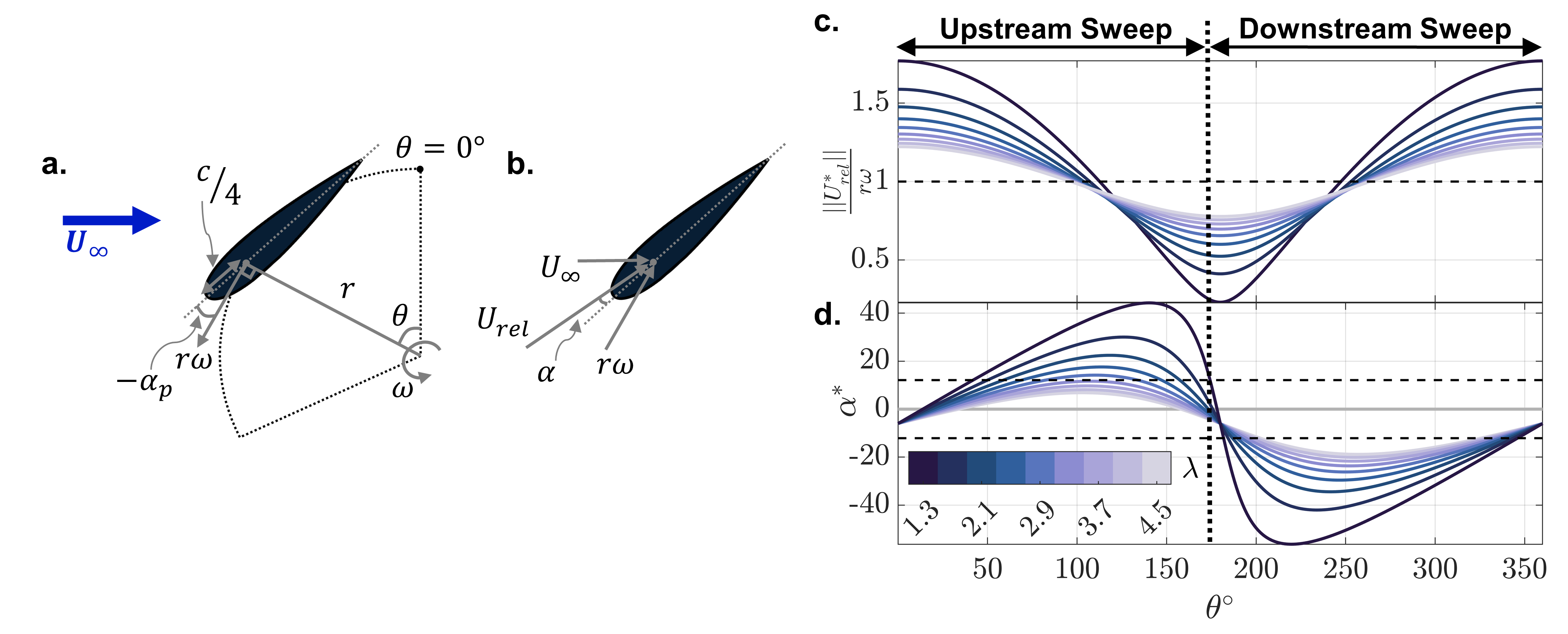}
    \caption{(a,b) Blade geometric definitions, (c) normalized nominal relative velocity trajectories, and (d) nominal angle of attack trajectories for a preset pitch angle, $\alpha_p$, of $-6^\circ$. The tangential velocity is defined as tangent to the circular blade path. A negative (i.e., ``toe-out'') preset pitch angle is depicted in (a), and (b) shows the angle of attack definition. The static stall angle ($\pm 12.1^\circ$, dashed lines) in (d) is for a foil in rectilinear flow at a similar Reynolds number\cite{staticalpha} ($Re_c\:=\:1.4x10^5$). Because of the rapidly varying angle of attack and appreciable induction, the comparison between $\alpha^*$ and the steady-state stall angle is qualitative.}
  \label{nominal}
\end{figure}

Overall, cross-flow turbine kinematics, dynamics, and performance for a specific geometry and set of inflow parameters are functions of the tip-speed ratio, $\lambda$, and the blade azimuthal position, $\theta$. Turbine geometric definitions are shown in figure \ref{nominal}. The tip-speed ratio is the non-dimensional ratio of the tangential velocity (product of turbine radius to the quarter chord, $r$, and rotation rate, $\omega$) to the freestream velocity, $U_\infty$, and is defined as
\begin{equation}
\lambda\:=\:\frac{r\omega}{U_{\infty}}.
\end{equation} 
We define a $0^\circ$ azimuthal position as corresponding to the location at which the blade tangential velocity vector points directly into the freestream. The upstream sweep spans $\theta\:=\:0^\circ\:-\:180^\circ$ and the downstream sweep spans $\theta\:=\:180^\circ\:-\:360^\circ$. During each rotation, blades encounter a continually fluctuating relative inflow velocity, $U_{rel}$, (affecting the magnitude of the fluid forces) and angle of attack, $\alpha$, (affecting lift, drag, and pitching moment coefficients). 
Descriptions of these kinematic terms often utilize ``nominal'' formulations ($U^*_{rel}$, $\alpha^*$) that assume the turbine does not affect the inflow and that velocities everywhere in the flow are equal to the free stream condition, $U_\infty$. 
The azimuthal variations in $||U^*_{rel}||$ and $\alpha^*$ over one rotation are shown in figure \ref{nominal}c,d. Here $||\cdot||$ denotes magnitude. The nominal incident velocity at the quarter chord, $c/4$, is the vector sum of the tangential and freestream velocities, such that its magnitude is 
\begin{equation}
||U^*_{rel}(\lambda,\theta)||=U_\infty\sqrt{\lambda^2+2\lambda cos(\theta)+1}.
\label{nominal Urel}
\end{equation}

\noindent The nominal angle of attack, defined as the angle between the chord line and $U^*_{rel}$ at $c/4$ is
\begin{equation}
\alpha^*(\lambda,\theta)=tan^{-1}\bigg[\frac{sin(\theta)}{\lambda+cos(\theta)}\bigg]+\alpha_p
\label{nominal aoa}
\end{equation}
where $\alpha_p$ is the blade preset pitch angle. As defined, a negative (i.e., ``toe-out'') preset pitch angle (figure \ref{nominal}a) decreases the maximum angle of attack magnitude in the upstream sweep but increases the 
maximum angle of attack magnitude in the downstream sweep. 
Nominally, a decrease in $\lambda$ reduces $||U^*_{rel}||$ and increases the range of $\alpha^*$ encountered during a cycle. Conversely, as $\lambda$ increases, $||U^*_{rel}||$ converges towards the tangential velocity (figure \ref{nominal}c).  

While neglected by the nominal kinematic description, the influence of the turbine on the inflow -- termed ``induction'' -- is substantial, particularly during the downstream sweep \cite{rezaeiha2018solidity,Li,Tescione2014}. 
The true relative velocity magnitude is the vector sum of the induction-modified inflow and the blade tangential velocity, while the true angle of attack depends on the angle between the relative velocity direction and the chord line. These vary with blade azimuthal position, so, in reality, blade kinematics differ appreciably between the upstream and downstream sweeps. 

The continual variation in the relative velocity and angle of attack experienced by cross-flow turbine blades results in phase-dependent power production and can lead to the unsteady, non-linear phenomenon of dynamic stall \cite{Me,Snortland2023,Mukul,Mukul2,sebstalldilema,BIANCHINI2016329,Simao,BUCHNER,Dunne}. Dynamic stall severity depends on the angle of attack range experienced by the blade. For relatively large angle of attack ranges (lower $\lambda$), the near-blade flow field in the upstream sweep is characterized by the formation, roll-up and shedding of an energetic dynamic stall vortex that is on the order of the blade chord. In contrast, vortex growth for relatively small angle of attack ranges (high $\lambda$), is comparatively limited \cite{Mulleners,McCroskey}. The topology of the near-blade flow structures differ substantially between the upstream and downstream sweeps, as would be expected from the differences in blade kinematics. 

Most research that has focused on interpreting and optimizing cross-flow turbine performance has primarily considered the power-generating phases of the upstream sweep, but a limited number of works discuss the hydrodynamics and/or performance differences between the upstream and downstream sweeps. Rezaeiha et al. \cite{REZAEIHA2018TSRimpact} showed computationally that for six tip-speed ratios, time-averaged performance contributions from the downstream sweep are substantially smaller than upstream, and that downstream performance degrades with increasing tip-speed ratio, while upstream performance improves. They explain the reduction in downstream performance with increasing tip-speed ratio as a consequence of greater momentum removal in the upstream sweep. 
Because the upstream sweep contributes the most to power production, they concluded it is most important for overall turbine performance, and, as such, should be the focus for power-enhancement strategies. However, Mulleners et al. \cite{Mulleners2023activepitch} experimentally applied an active pitch controller and found performance at a high tip-speed ratio could be improved predominately by changes in power production in the downstream sweep. Recently, Le Fouest et al. \cite{sebtimescales} experimentally investigated the time-scales associated with dynamic stall and cross-flow turbine forcing. They attributed changes in the hydrodynamics and performance between the upstream and downstream to differences in dynamic stall development stemming from the asymmetry of the nominal angle of attack trajectories between the upstream and downstream sweeps. However, their description excludes induction which modifies the nominal kinematics. 

In addition, several groups have attempted to explain performance variations for different turbine geometries through observed changes in the upstream and downstream regions. Hunt et al. experimentally \cite{hunt2023parametric} and Rezaeiaha et al. computationally \cite{REZAEIHA2017presetpitch} highlight opposing trends in performance between the upstream and downstream sweeps as a function of the preset pitch angle. Hunt et al. show that the present pitch angle that is optimal for overall turbine performance represents a compromise between improvements to the upstream sweep and degradation in the downstream sweep. Li et al. \cite{LI2016solidity} discuss how force coefficients from experimental surface pressure measurements differ between the upstream and downstream sweeps for different turbine solidities (
varied by changing the blade count, $N$). 
They hypothesize that pressure differences across the blade decrease in the downstream sweep because induction reduces the magnitudes of the angle of attack and relative velocity. Similarly, Rezaeiha et al. \cite{rezaeiha2018solidity} computationally highlight opposing upstream/downstream performance trends with solidity (varied by changing $c/r$ and $N$). As in their other work \cite{REZAEIHA2018TSRimpact}, decreased power generation in the downstream sweep at higher solidities is attributed to reductions in available momentum.

These prior works discuss different trends between the upstream and downstream sweeps, suggest linkages between them, and pose three explanatory mechanisms for downstream performance degradation relative to upstream:
(1) reduced momentum available in the downstream sweep \cite{REZAEIHA2018TSRimpact,rezaeiha2018solidity},
(2) different blade kinematics (angle of attack, relative velocity) throughout the downstream sweep \cite{LI2016solidity}, and
(3) the near-blade dynamics in the downstream sweep associated with dynamic stall and flow recovery, interactions with previously shed coherent structures, and the reversal of the pressure and suction sides of the blade \cite{sebtimescales}. 
All three mechanisms are inextricably linked through induction, but prior work has not described the relationships between them or addressed the relative importance of these mechanisms. 

Here, experimental turbine performance is measured in concert with phase-locked particle image velocimetry (PIV) flow fields for a one-bladed turbine to directly characterize the dynamics and performance contributions of the downstream sweep. 
The paper is laid out as follows. Section~\ref{methods} presents the methodology for the turbine performance and flow-field measurements. Section~\ref{results} describes how performance, induction, and the near-blade dynamics differ between the upstream and downstream sweeps. Then, the explanatory mechanisms responsible for downstream performance degradation are explored, and the accuracy of the nominal formulations of the angle of attack and relative velocity in the downstream sweep are evaluated. The effect of induction on the kinematics and dynamics of cross-flow turbines is explored, and a new analytical model for a limiting case where in-rotor velocities are negligible is presented. In section~\ref{discussion}, the observed downstream performance trends are shown to generalize through reanalysis of 54 additional unique performance experiments from \cite{hunt2023parametric} that parametrically vary preset pitch angle, chord-to-radius ratio, and the Reynolds number.
   
\section{Methods}
\label{methods}
We explore the influence of the downstream blade sweep on cross-flow turbine performance using a one-bladed turbine. Since, in our experiments, turbine torque is measured at the center shaft, a one-bladed turbine allows us to isolate the performance contributions for the upstream and downstream portion of the rotation (i.e., with a multi-bladed turbine, the torque contribution from each blade is ambiguous). Flow fields are investigated to contextualize the hydrodynamics that underly the observed performance trends. For this purpose, two-component, phase-locked, planar PIV data is obtained inside the turbine swept area at the blade mid-span for three tip-speed ratios, $\lambda\:=\:1.4,2.4,$ and $3.4$.

\subsection{Experimental Facility}
\label{flumedescript}
\begin{figure*}[t!]
    \centering
    \includegraphics[width=1\linewidth]{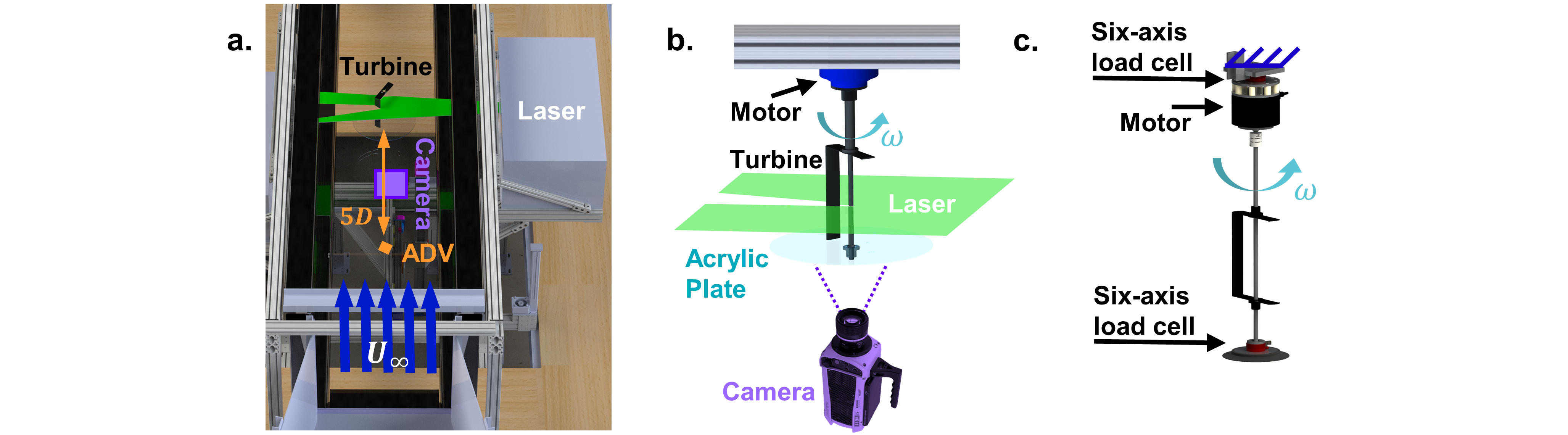}
    \caption{(a) Annotated PIV and performance experimental setup in the flume, (b) ``PIV measurement" turbine setup with the camera and laser sheet arrangement, and (c) ``performance measurement" turbine setup.}
    \label{flume}
\end{figure*}
Experiments were performed in the Alice C. Tyler flume at the University of Washington (figure \ref{flume}a). The data presented in this paper utilized a mean dynamic water depth, $h$, of 0.52 m, resulting in a channel cross-sectional area $A_C$ of 0.39 m\textsuperscript{2} (0.75 m width). The water temperature was maintained at $39\pm0.2$ \textsuperscript{$\circ$}C, giving a density, $\rho$, of 993 kg/m\textsuperscript{3}, and a kinematic viscosity, $\nu$, of $6.7\times10^{-7}$ m\textsuperscript{2}/s. An acoustic Doppler velocimeter (Nortek Vectrino) positioned approximately 5 diameters upstream of the turbine rotor measured the inflow at a 16 Hz sampling rate. After de-spiking \cite{GoringandNikora}, the average $U_{\infty}$ was 0.9 m/s with a turbulence intensity of 1-2\%. These conditions corresponded to a depth-based Froude number, $Fr\:=\:U_\infty/\sqrt{gh}$, of 0.4 where $g$ is the gravitational constant. 

\subsection{Cross-flow Turbines and Performance Characterization}
\label{test rigs}
Two experimental turbine setups were used in these experiments: the ``PIV measurement'' setup (figure \ref{flume}b) and the ``performance measurement'' setup (figure \ref{flume}c). In either configuration, turbine rotation rate was regulated by a servomotor to achieve a desired $\lambda$. Blade position was measured by the servomotor encoder with a resolution of $2^{18}$ counts/rotation and $\omega$ was computed by differentiation. MATLAB Simulink Desktop Real-Time was used for data collection and turbine control. For each control set point (i.e., a single $\lambda$), all data were acquired for $60$ seconds at 1 kHz. The turbine had a radius of 8.2 cm to the quarter-chord mounting location, a diameter, $D$, between the blade outer surfaces of 17.2 cm, a blade span, $H$, of 23.4 cm, a blade chord length of 4.06 cm, a preset pitch angle of $-6^\circ$, and a NACA 0018 profile. The turbine solidity,
\begin{equation}
\sigma\:=\:\frac{Nc}{2\pi r},
\label{solidity}
\end{equation}
was $0.078$. The blockage ratio, $\beta\:=\:\frac{HD}{A_C}$, was 10.3\% and the Reynolds number,  $Re_c\:=\: \frac{U_{\infty}c}{\nu}$, was $5.5\times10^4$. 

The ``PIV measurement'' test setup utilized a servomotor (Yaskawa SGMCS-02B) rigidly coupled to the flume cross beam and controlled with a drive (Yaskawa SGDV-2R1F01A002). To facilitate PIV imaging of a streamwise plane at the blade mid-span using a camera positioned below the flume, the turbine was constructed with a 40 cm diameter (2.3x turbine diameter) acrylic plate at the bottom and a NACA 0008 foil strut at the top. The plate was intentionally oversized so that the index of refraction was constant across the field of view. This turbine, however, was sub-optimal for performance measurements due to high drag on the acrylic plate (particularly at high rotation rates) and forces and torques could not be measured with this test setup. 

In the ``performance measurement'' test setup, forces and torques were measured by a pair of six-axis reaction load cells (above rotor: ATI Mini45, below rotor: ATI Mini40). This setup used a different servomotor (Yaskawa SGMCS-05B) but the same drive as for the ``PIV measurement'' setup. For more realistic performance, this turbine was comprised of NACA 0008 foil struts supporting each end of the blade span. Since the struts still incur an appreciable parasitic torque, blade-level performance was isolated by subtracting phase-averaged performance for the turbine support structure (blade removed, same operating conditions) from the full turbine performance. We note that this relies on an assumption that secondary interactions between the blades and support structures are minimal, which is supported by prior work \cite{stromsupports,hunt}. All torque measurements were filtered with a low-pass, zero-phase, Butterworth filter to remove high-frequency electromagnetic interference from the servomotor. The 30 Hz cutoff frequencies used for the turbine and support structure performance data are approximately $10$ harmonics faster than the blade passage frequency, so the filter is unlikely to remove any hydrodynamic torque.

Over the course of a turbine rotation, the phase-varying hydrodynamic power, $P$, was non-dimensionalized as the coefficient of performance, $C_P$, defined as,
\begin{equation}
  C_P(\lambda,\theta)\:=\:\frac{P(\theta)}{\frac{1}{2}\rho U^3_\infty DH},
\label{CP}
\end{equation}
where $\rho$ is fluid density. Throughout, we refer to the coefficient of performance with the shorthand of ``performance''.
The hydrodynamic torque, $Q$, was non-dimensionalized as 
\begin{equation}
C_{Q}(\lambda,\theta)\:=\:\frac{Q(\theta)}{\frac{1}{2}\rho U^2_\infty DHr},
\label{CQ}
\end{equation} 
noting that the torque and performance coefficients are directly related by the tip-speed ratio as
\begin{equation}
C_P\:=\:C_{Q} \lambda.
\label{torque perf relation}
\end{equation}

Time-averages of any quantity, $X$, were calculated over an integer number of rotations for a single $\lambda$ set point and represented as $\overline{X}$. We also present averages that are conditional on $\theta$ segments. For example, the upstream segment-averaged performance coefficient was computed by averaging all of the data points in the time-series for $0^\circ\:\leq\:\theta\:<\:180^\circ$ . A corresponding calculation was used to define a segment-averaged downstream performance coefficient. In this work, the upstream and downstream segment averages are scaled by 1/2 such that their sum is equal to the time-averaged performance for a full rotation. Phase averages of any quantity, $X$, for a single azimuthal position across multiple cycles at a $\lambda$ set point are represented as $\langle X \rangle$. Since the performance data is captured continuously, we utilized an azimuthal bin of $1^\circ$ for phase-averaging.

\subsection{PIV Measurement}
Two-dimensional, two-component, phase-locked flow-field measurements were obtained in a streamwise plane at the blade mid-span. PIV acquisition was controlled by TSI Insight 4G (version 11.0.1) and acquisition for each cycle commenced upon receipt of trigger pulses sent at a specified $\theta$ from the Simulink model controlling the turbine. 

The general arrangement of the PIV laser and cameras is shown in figure \ref{flume}a. A dual cavity, Nd:YLF laser (Continuum Terra PIV) illuminated the flow with an approximately 2 mm thick horizontal light sheet in the cross-stream direction. A high-speed camera  (Vision Research Phantom v641) with 2560 x 1600 resolution acquired images from below. The flow seeding (10 $\mu$m hollow-glass beads) produced particle images of approximately 3 pixels in diameter. The PIV data were taken during two different experiments. The first experiment captured flow field data for $\lambda\:=\:1.4$ and $\lambda\:=\:2.4$, and utilized a 60 mm lens at f\#16 resulting in a calibration of 7.9 pixels/mm and a field of view, FoV, of 32.4 x 20.3 cm [$8c$ ($1.9D$) x $5c$ ($1.2D$)]. The second experiment captured flow field data for $\lambda\:=\:3.4$ and utilized a 50 mm lens at f\#4 resulting in a calibration of 8.22 pixels/mm and 31.1 x 19.5 cm [$7.6c$ ($1.8D$) x $4.8c$ ($1.1D$)] FoV. The limited streamwise extent of the laser sheet necessitated shifting the turbine by $\approx\frac{1}{2}D$ upstream to capture the downstream blade sweep (logistically preferred to shifting the laser). Cross-stream FoV positioning relied on camera movement with a motorized, three-axis gantry which also allowed for fine adjustments in the streamwise direction.

Sequences of 59 image pairs were acquired per rotational cycle with prescribed angular displacements of approximately $3^\circ$ between frames for $\theta\:=\:9^\circ\:-\:176^\circ$ and $\theta\:=\:183^\circ\:-\:353^\circ$. Image pairs were captured at each phase over twenty rotations. Both the turbine shaft and blade cast shadows in the laser sheet. Therefore, to obtain data adjacent to the suction and/or pressure sides of the foil, PIV measurements were conducted with the turbine spinning in both clockwise and counter-clockwise directions, with vector fields combined in post-processing. To minimize laser reflections at the blade surface, matte black paint was applied. 

PIV processing was performed in LaVision DaVis (version 10.2.1). Background subtraction using a Butterworth filter on phase-matched images mitigated residual reflections and background illumination variation. The shadowed regions, visible turbine structures, and remaining reflections were manually masked. The applied mask is unique to each azimuthal position but constant for all rotational cycles. Processing utilized a multi-grid, multi-pass cross-correlation algorithm with adaptive image deformation, resulting in final window sizes of 32 x 32 pixels with a 75\% overlap ($\sim$40 vectors per chord length). Spurious vectors were removed with a universal outlier median filter \cite{outlierdetect} utilizing a 9x9 filter region and a threshold value of 1.5 for the $\lambda\:=\:1.4$ case and 2.5 for the $\lambda\:=\:2.4$ and 3.4 cases.    

\subsection{Flow Field Analysis}
Flow fields from both rotation directions were combined in post-processing in MATLAB. To produce the composite fields, the center of rotation was located and aligned between the different FoVs. Tracking a small dot on the end of the shaft and registering all resultant PIV vector fields to a common shaft location corrected phase-dependent shaft procession (slight run-out on the cantilevered turbine shaft). Because of parallax and differences in the index of refraction between air and water, the imaged center of the turbine shaft does not correspond to the center of rotation of the turbine at the imaging plane. Therefore, we determined a best-fit location of the center of rotation in each FoV by manually fitting blades to the masked regions. The phase-averaged flow fields were then interpolated to a common grid and the two rotation directions at each phase with the same FoV were averaged. 

The flow field analysis utilized segment-averaged velocity magnitude fields, $||\overline{\vec{V}}||$ in a flume reference frame, and phase-averaged relative velocity fields in the blade-centric reference frame, $||\langle \vec{U_{rel}} \rangle||$. The $||\overline{\vec{V}}||$ fields were computed separately for the upstream and downstream sweeps and are an average of the composite flow fields at all blade positions. Any regions of data missing in $\geq20\%$ of the composite flow fields (regions that are in the masked areas in both rotation directions) were ignored in the segment-average. The flow fields were translated to the blade-centric reference frame by first rotating the entire field about the center of rotation to a common blade orientation and then computing the $||\langle \vec{U_{rel}}\rangle||$ fields as the vector sum of the flume-reference velocity fields and the blade tangential velocity. 

\begin{figure}[t!]
    \centering
    \includegraphics[width=1\linewidth]{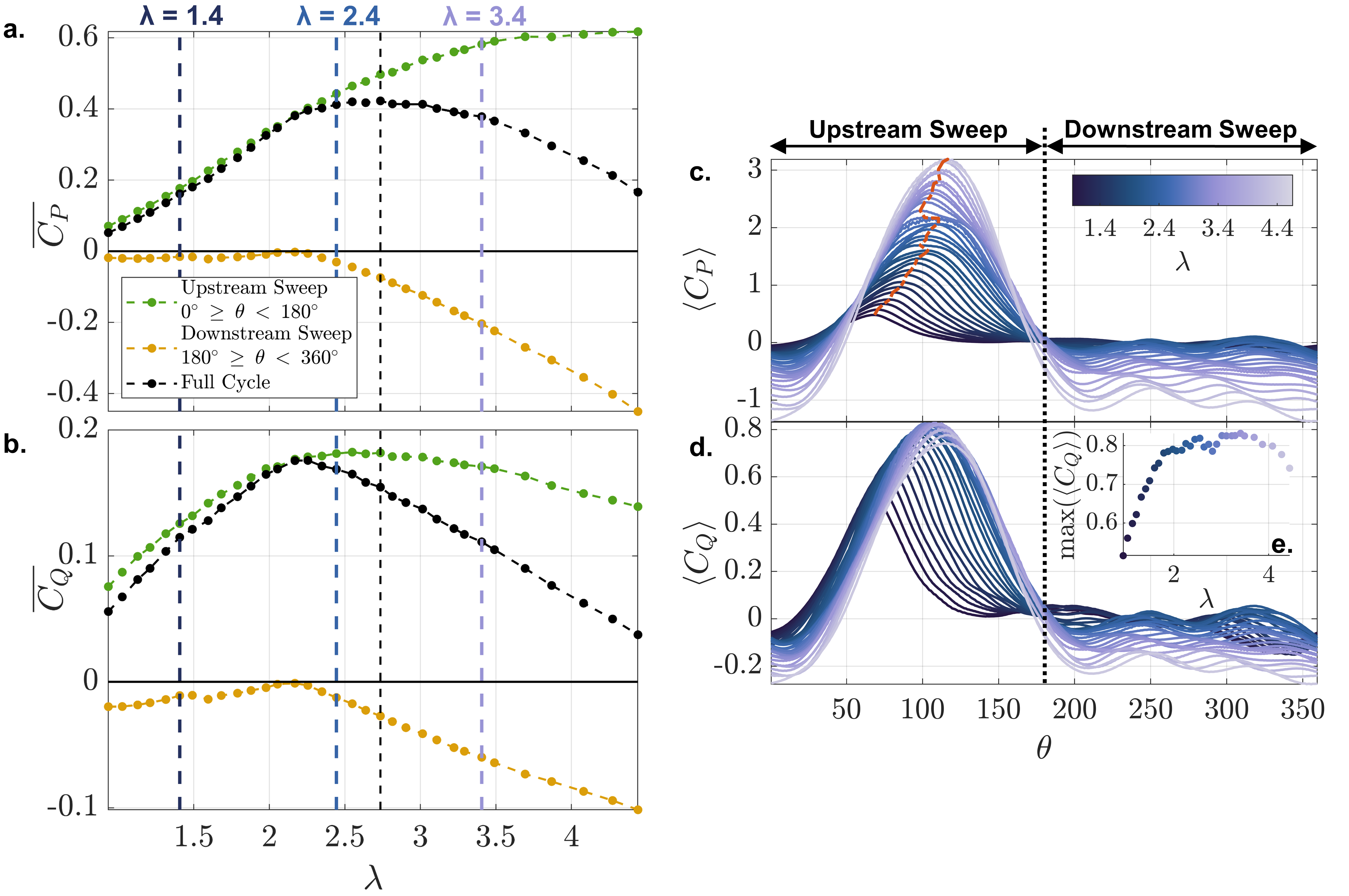}
    \caption{(a) Time- and segment-averaged coefficient of performance and (b) coefficient of torque. The upstream and downstream segment-averages are scaled by 1/2 such that their sum is equal to the total time-averaged performance. The vertical, black dashed line corresponds to the optimal tip-speed ratio and the colored, dashed lines denote the tip-speed ratios where flow fields were acquired. (c) Phase-averaged performance coefficient and (d) torque coefficient. (e) The value of the maximum torque coefficient as a function of tip-speed ratio. The orange dashed line in the upstream sweep in (c) tracks the phase of maximum performance.} 
    \label{torque and perf}
\end{figure}

\begin{figure}[t!]
    \centering
    \includegraphics[width=0.49\linewidth]{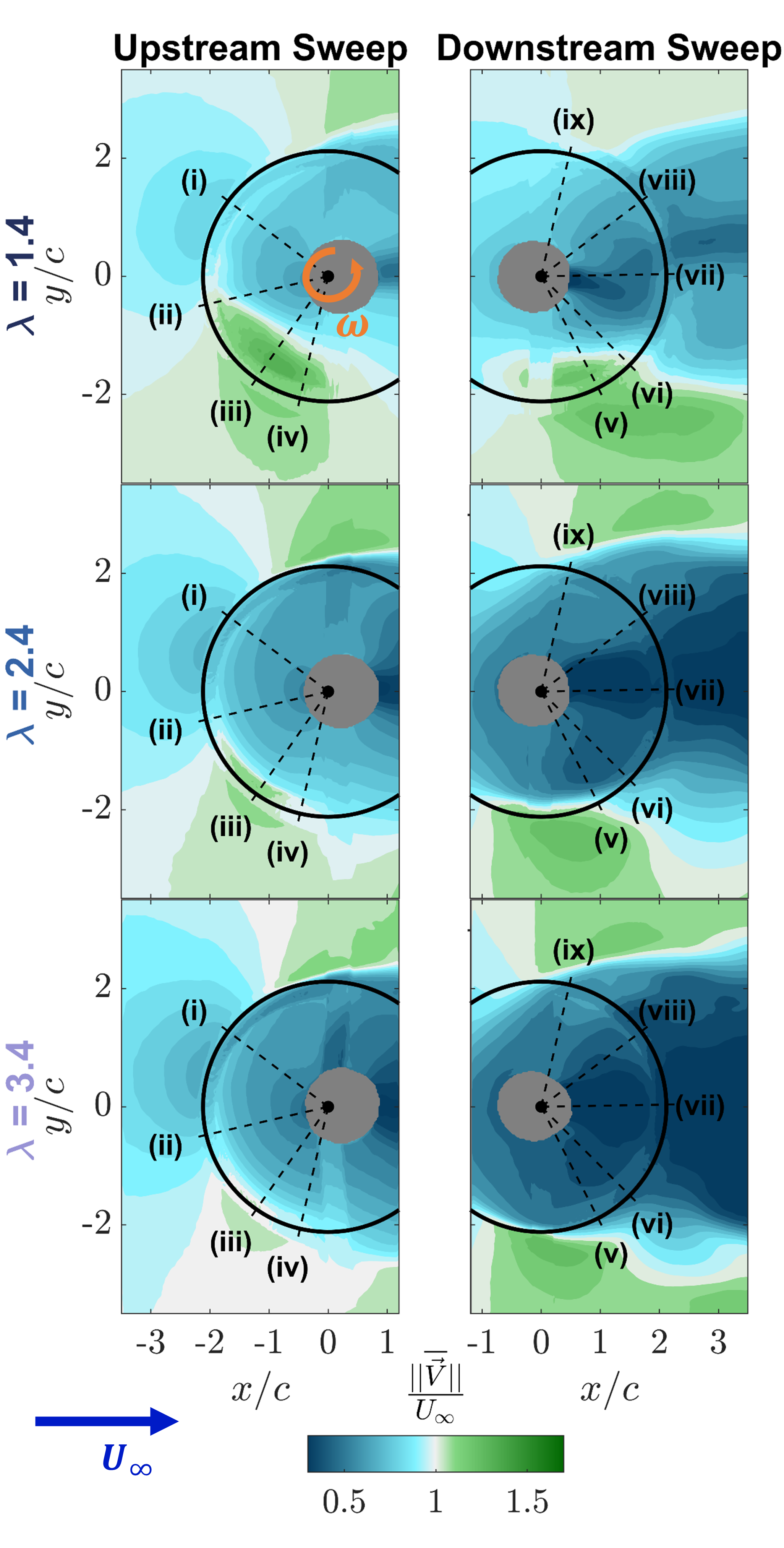}
    \caption{Segment-averaged horizontal velocity magnitude fields for the upstream and downstream sweeps for $\lambda\:=\:1.4$, $\lambda\:=\:2.4$, and $\lambda\:=\:3.4$. Freestream flow is from left to right. The colorbar has been truncated at 0.3 and 1.7 for visualization. The black circles represent the blade sweep. The radial dashed lines correspond to the locations of the phase-averaged flow fields presented in figure \ref{flow phase}c,d. The upstream sweep is averaged over $\theta\:=\:9^\circ\:-\:176^\circ$ while the downstream sweep is averaged over $\theta\:=\:183^\circ\:-\:353^\circ$. The dark gray areas represent the masked regions.}
    \label{flow time-average}
\end{figure}

\section{Results}
\label{results}
\subsection{Performance and Torque Measurements}
As shown in figure \ref{torque and perf}a, time-averaged turbine performance increases with tip-speed ratio up to an optimal value ($\lambda_{opt}\:=\:2.7$), beyond which performance begins to decrease. However, when the upstream and downstream contributions to this time-averaged performance are partitioned, it is clear that upstream sweep performance continues to increase beyond the optimal tip-speed ratio. In contrast, downstream sweep performance is approximately net neutral (i.e., segment-average $C_P \approx 0$) until $\lambda \approx 2.4$, after which it begins to decrease at a faster rate than the upstream performance increases. This indicates that the optimal tip-speed ratio for this turbine is strongly influenced by the onset of continuous performance degradation in the downstream sweep.

Figure \ref{torque and perf}b shows that the time-averaged torque coefficient in the upstream sweep slowly declines beyond the peak tip-speed ratio. Since turbine performance is the product of the torque coefficient and the tip-speed ratio (equation \ref{torque perf relation}), this indicates that the tip-speed ratio drives the upstream performance increase beyond $\lambda_{opt}$. Therefore, not only does the tip-speed ratio influence the kinematics, near-blade dynamics, and torque production, but the tip-speed ratio can compensate for moderate torque losses in the upstream sweep in terms of performance. In the downstream sweep, torque is increasingly negative, so the tip-speed ratio only exacerbates detrimental downstream performance (i.e., power consumption).
Additional support for this conclusion is provided by the phase-averaged coefficients of performance and torque (figure \ref{torque and perf}c-e). In the upstream sweep, the amplitude of the performance peak increases continuously with the tip-speed ratio (figure \ref{torque and perf}c). As the tip-speed ratio increases, the upstream torque peak grows in width and amplitude and shifts to later in the cycle until the timing and amplitude of the peak become relatively independent of $\lambda$ for $\lambda\:=\:2\:-\:4$, beyond which peak torque begins to decrease (figure \ref{torque and perf}d,e). In contrast to the upstream sweep, performance and torque in the downstream sweep (figure \ref{torque and perf}c,d) are relatively independent of phase and increasingly detrimental as the tip-speed ratio increases. For all experiments, the parasitic performance at the end of the downstream sweep persists into the beginning of the upstream sweep.



\begin{figure}[t!]
    \centering
    \includegraphics[width=0.93\linewidth]{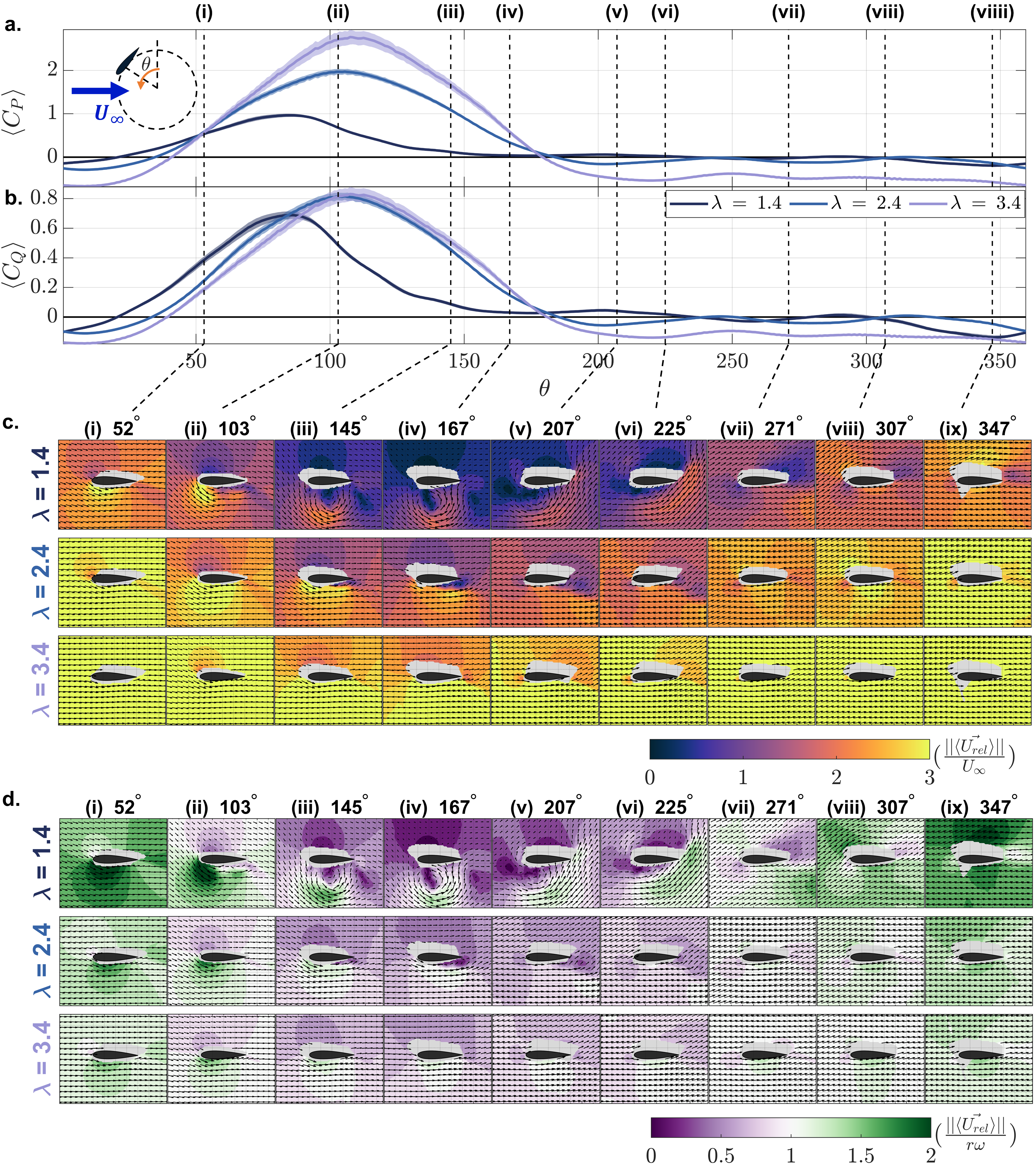}
    \caption{(a) Select phase-averaged performance (b) and torque corresponding to tip-speed ratios with PIV data ($\lambda\:=\:1.4$, $\lambda\:=\:2.4$, and $\lambda\:=\:3.4$). The shaded regions represent $\pm1$ standard deviation across phase-averages of individual cycles. The vertical dashed lines correspond to the locations of the phase-averaged flow fields presented in figure \ref{flow phase}c,d. Phase-averaged relative velocity fields normalized by (c) the freestream velocity and (d) the tangential velocity at these phases. The velocity vectors are downsampled for visualization and the grid spacing is $c/4$. There are no vectors in the masked regions.}
    \label{flow phase}
\end{figure}





While the significant influence of the downstream sweep is clear, these data cannot entirely identify the mechanisms for the continuous downstream performance degradation. As previously discussed, three explanatory mechanisms have been proposed in prior work: reductions in available momentum, changes to blade kinematics, and changes to the near-blade dynamics. To evaluate the relative importance of these mechanisms, we explore the extent of induction and the near-blade flow structures. Turbine flow fields are analyzed for three contrasting performance cases (one near-optimal, $\lambda\:=\:2.4$, and two sub-optimal, $\lambda\:=\:1.4$ and 3.4). The $\lambda\:=\:2.4$ case, which is slightly less than optimal ($\lambda\:=\:2.7$), corresponds to the point where the downstream performance switches from being consistently nearly net-neutral to decreasing monotonically. 

The segment-averaged in-rotor velocity is shown in figure \ref{flow time-average} and seen to be slower than the freestream within the majority of the rotor area. The in-rotor velocities generally decrease as the tip-speed ratio increases while, in all cases, there are isolated pockets of the upstream region of the turbine swept area with velocities higher than $U_\infty$. We also observe velocities exceeding $U_\infty$ in the bypass region between the turbine and the flume walls as a result of the divergence of the flow around the turbine and confinement. These results are generally consistent with linear momentum actuator disc theory for confined flow which predicts that flow should decelerate through the rotor but accelerate around it \cite{GARRETT_CUMMINS_2007}. The most pronounced deceleration occurs in the downstream portions of the rotor and in the wake, where velocities are as low as 8\% of $U_\infty$ for the highest tip-speed ratio. This demonstrates greater momentum loss through the turbine at higher tips-speed ratios and is consistent with Rezaeiha et al.'s \cite{REZAEIHA2018TSRimpact,rezaeiha2018solidity} conclusions from computational studies.  

Significant changes to the free stream velocity from induction degrade the accuracy of the nominal descriptions for the relative velocity (equation~\ref{nominal Urel}) and angle of attack (equation~\ref{nominal aoa}) in the upstream and downstream sweeps. To evaluate the actual turbine kinematics, we consider select phase-averaged, blade-centric flow fields normalized by the freestream velocity (figure \ref{flow phase}c) and by the tangential velocity (figure \ref{flow phase}d). Corresponding phase-averaged performance and torque (a subset of the tip-speed ratios in figure \ref{torque and perf}) are given in figure \ref{flow phase}a and figure \ref{flow phase}b, respectively. As with performance and torque, the flow fields depend strongly on $\lambda$ and $\theta$, differing substantially between the upstream and downstream sweeps. Despite significantly lower in-rotor velocities at high tip-speed ratios (figure \ref{flow time-average}), blade-relative velocities increase with tip-speed ratio at all phases (i.e., orange and yellow hues in figure \ref{flow phase}c) and converge towards the tangential velocity (i.e., lighter hues in figure \ref{flow phase}d), particularly in the downstream sweep. 
In other words, as $\lambda$ increases, the higher tangential blade velocities outweigh the near-blade inflow velocities (which are reduced by increasing induction), such that the relative velocity field becomes dominated by the tangential velocity. While this also clearly affects the angle of attack, the heterogeneous nature of the relative velocity field defies a consistent definition for this kinematic term. 

We now consider the influence of the near-blade dynamics. Overall, the complexity of the near-blade dynamics and the tip-speed ratio are inversely proportional for the three cases, as apparent by the weakening coherent structures throughout the turbine rotation at higher tip-speed ratios. This is in qualitative agreement with prior experiments \cite{sebstalldilema,Simao,Me,Snortland2023} and simulations \cite{Coriolis,Mukul,REZAEIHA2018TSRimpact}. The $\lambda\:=\:1.4$ case has the most complex near-blade dynamics which include the roll-up and shedding (figure \ref{flow phase}c: ii-iv) of a strong dynamic stall vortex in the upstream sweep, as well as prolonged post-stall flow separation (v-vi) and more persistent separated flow (vii-ix) during the downstream sweep. The relatively high velocity region between $\theta\:=\:90^\circ-180^\circ$ for $\lambda\:=\:1.4$ (figure \ref{flow time-average}) is a signature of the strong dynamic stall vortex. In contrast, higher tip-speed ratios have simpler near-blade dynamics, as evidenced by smaller vortex structures, limited flow separation (i-iv), faster post-stall flow recovery (v-vi), and minimal separated flow (vii-ix) in the downstream sweep. 

The observed variation of the kinematics and near-blade dynamics qualitatively explain the increased upstream torque production for $\lambda\:=\:2.4$ and $\lambda\:=\:3.4$ as compared to $\lambda\:=\:1.4$. As the tip-speed ratio increases, dynamic stall weakens, stall onset is delayed to later in the cycle, and the relative velocity incident on the blade increases (figure \ref{flow phase}c). In aggregate, these mechanisms increase the amplitude of the torque peak and shift the peak later in the cycle until the phase of maximum torque becomes nearly independent of $\theta$ (figure \ref{torque and perf}c). Similar trends are observed in the downstream sweep, where the phase-averaged flow fields indicate more attached flow and higher relative velocities as the tip-speed ratio increases. A trend towards attached flow would be expected to decrease the drag coefficient and increase lift, but, contrary to the upstream sweep, downstream torque production and performance are initially near zero, then decrease linearly with $\lambda$ (figure \ref{torque and perf}a,b). 
Notably, the downstream near-blade dynamics differ the most between $\lambda\:=\:1.4$ and $\lambda\:=\:2.4$, but the corresponding differences in downstream segment-averaged torque and performance are small. This is as a consequence of opposing torque fluctuations in the phase-averages (figure \ref{flow phase}b). 
In contrast, while the near-blade dynamics are relatively unchanged between $\lambda\:=\:2.4$ and $\lambda\:=\:3.4$, downstream performance and torque are significantly more negative for $\lambda\:=\:3.4$. This mismatch in performance and flow field trends indicates the near-blade dynamics are unlikely to explain the downstream performance degradation.

\begin{figure}[t!]
    \centering
    \includegraphics[width=1\linewidth]{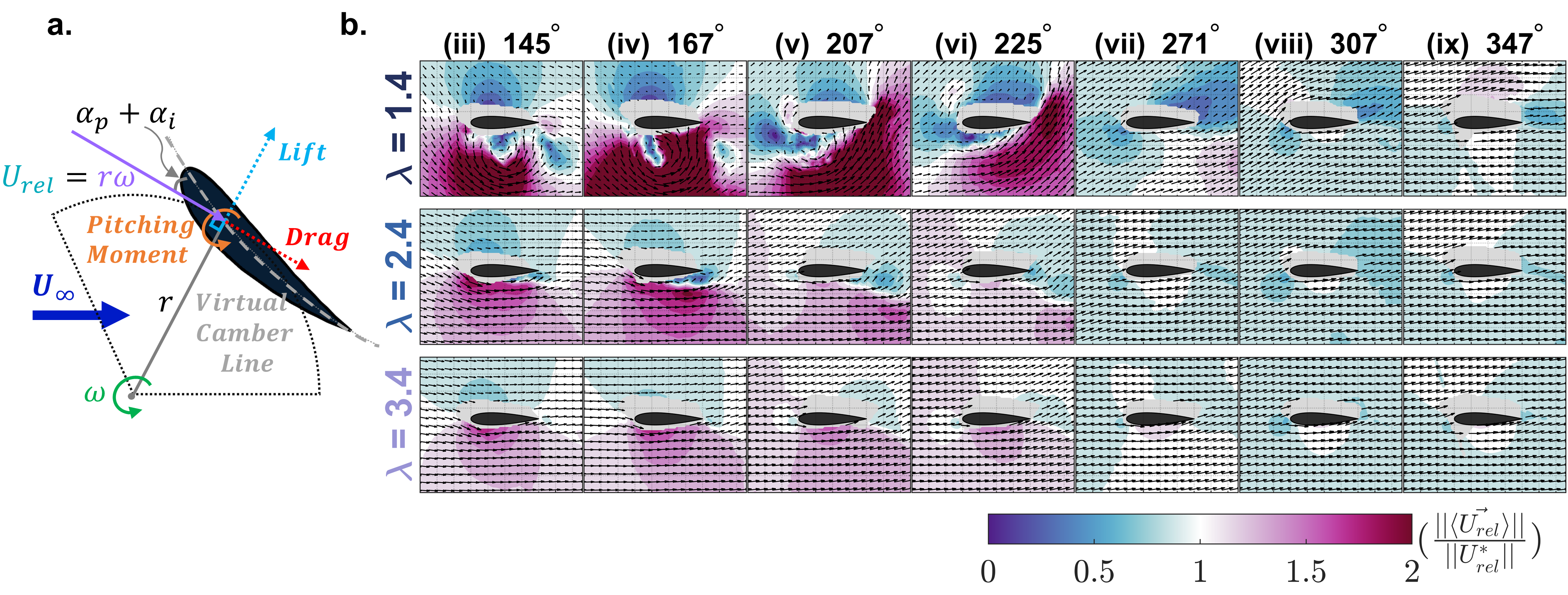}
    \caption{(a) Force and kinematic schematic for the limiting case where the relative velocity is equal to the blade tangential velocity. The blade shape reflects an exaggeration of virtual camber and neither the blade chord or radius are to scale. (b) Phase-averaged relative velocity fields normalized by the nominal relative velocity (equation \ref{nominal Urel}). The velocity vectors are downsampled for visualization and the grid spacing is $c/4$. The masked regions have no vectors.}
    \label{lift drag}
\end{figure}

Instead, the increasingly detrimental performance in the downstream sweep is primarily a consequence of reductions to available momentum and changes to the kinematics (i.e., a combination of two proposed mechanisms). We note that even in the absence of induction, blade-relative velocities will converge to the tangential velocity at high tip-speed ratios. This suggests the existence of a limiting case at a sufficiently high tip-speed ratio, $\lambda_{lim}$, for which $U_{rel}\:=r\omega\:$. Cross-flow turbine performance depends, at the blade level, on the tangential projection of the resultant force (combination of lift and drag acting at $c/4$) in the direction of rotation, pitching moment (arising from a center of pressure displaced from $c/4$), rotation rate, and turbine radius to $c/4$.
At $\lambda_{lim}$, the lift is perpendicular to the rotation direction (figure \ref{lift drag}a), so has no tangential component and is unable to affect torque. In contrast, drag always opposes the rotation direction and increases with the tangential velocity. As the tip-speed ratio and induction increase, the in-rotor velocities decrease, and more of the downstream sweep approaches the limiting case (figure \ref{flow phase}d). In summary, at high tip-speed ratios, the orientation of lift and drag are unfavorable for power generation in the downstream sweep and detrimental drag is magnified by higher relative velocities. Therefore, induction limits power production in the downstream sweep not only because less momentum is available, but lift and drag act in less favorable directions. The pitching moment contribution will either enhance or detract from downstream performance, but further work is necessary to quantify its effect. 


Finally, we emphasize that the nominal model (equations \ref{nominal Urel} and \ref{nominal aoa}) poorly describes the kinematics throughout the blade rotation. Figure \ref{lift drag}b highlights how the measured relative velocities differ from the nominal model, as well as the challenges of computing representative values for true relative velocity and angle of attack due to strong spatial variation. In contrast to the hypothesis by Li et al. \cite{LI2016solidity} that the relative velocity magnitude is reduced in the wake, the relative velocities incident on the leading edge of the blade in the low-momentum region (vi-viii) are not always smaller than the nominal predictions. Notably, between the pressure and suction sides of the blade, opposing differences with respect to the nominal values could produce a spatial average that roughly matches the nominal model.

\subsection{Modeling the Downstream Sweep}
\label{Downstream Model}

Given the inability of the nominal kinematic model to describe the downstream sweep, we introduce an alternative analytical model for the limiting case (based on that derived for the no-inflow condition in \cite{MiglioreWolfe}) for downstream performance at high tip-speed ratios. Under these conditions $U_{rel}$ and $\alpha$ are invariant to $\theta$ (steady-state aerodynamics). 
However, even for the limiting case, cross-flow turbines are subject to rotational effects known as virtual camber, $C_v$, and virtual incidence (change in the perceived angle of attack due to flow curvature), such that a physically symmetric foil behaves like a cambered one \cite{Coriolis,MiglioreWolfe} and the pitching moment is non-zero. 
At the limiting case, virtual incidence and camber are functions $c/r$ and $\alpha_p$, and the foil is virtually cambered as in figure \ref{lift drag}a. Consequently, the aerodynamic coefficients are functions of the preset pitch angle and virtual geometric changes (camber and incidence) caused by rotation. Depending on the virtual camber and the preset pitch angle, during the downstream sweep, the lift vector could act towards the axis of rotation even for slightly negative effective angles of attack (the sum of the angle of attack and the virtual incidence).

The limiting case model is defined as follows: $U_{rel}$ is equal to $r\omega$, $\alpha$ is equal to the sum of $\alpha_p$ and the virtual incidence, and all steady-state aerodynamic coefficients are determined based on the virtually cambered foil shape. Under these conditions, cross-flow turbine torque, $Q_{lim}$, for a given $c/r$ and $\alpha_p$ is a function of the drag coefficient, $C_D(C_v,\alpha)$, pitching moment coefficient, $C_M(C_v,\alpha)$, and tangential velocity as
\begin{equation}    
    Q_{lim}=C_D \frac{1}{2}\rho (r \omega)^2 Hcr + C_M \frac{1}{2}\rho (r \omega)^2 Hc^2 .
    \label{torque}
\end{equation}
The corresponding coefficient of performance, $C_{P,lim}$ is (full derivation in supplemental information)

\begin{equation}    
    C_{P,lim}=\frac{r}{D}\Biggl[-C_D  \frac{c}{r} + C_M  \Bigl(\frac{c}{ r}\Bigr)^2\Biggr]\lambda^3 .
    \label{CPlimit}
\end{equation}
We note that $r/D$ is not identically 1/2 because the radius is defined relative to the quarter chord and diameter is defined relative to the blade surface. The drag term is always detrimental to performance (negative torque) while the pitching moment term may either contribute or hinder performance depending on its sign (positive corresponding to a pitch-in -- towards the center of rotation -- moment in the direction of blade rotation, figure \ref{lift drag}a). We can make some estimates about the relative importance of the terms in equation \ref{CPlimit} using NACA 6418 airfoil data. This foil has a similar profile to virtual camber estimated for our geometry at the limiting case: maximum camber of $\approx6\%$ occurring between the quarter and half chord locations (following the zero inflow formulation from \cite{MiglioreWolfe}). The corresponding virtual incidence is $\approx 6^\circ$, resulting in a $0^\circ$ angle of attack. Under these assumptions, $C_D\approx0.02$ and the pitching moment would be detrimental to performance, $C_M\approx-0.1$. For $c/r=0.49$, the drag-related term is approximately -0.01, and the pitching moment term is approximately -0.024, suggesting that the influence of virtual camber and pitching moment can be significant and detrimental to performance in the downstream sweep at the limiting case. We emphasize that this is only an order of magnitude analysis and a more complete evaluation would be beneficial. Preliminary calculations suggest that virtual camber may result in foils that depart from conventional NACA definitions in a number of ways, resulting in pressure gradients that may lead to transition or Reynolds number sensitivity, as well as relatively large changes in drag and pitching moment.

Finally, we observe that this model predicts that downstream performance should scale with $\lambda^3$. Instead, we observe an approximately linear relationship between $\lambda$ and the performance degradation in the downstream sweep beyond $\lambda\:=\:2.4$ (figure \ref{torque and perf}a). This suggests that $\lambda_{lim}$ is beyond the tip-speed ratio range tested for this turbine geometry. This is not entirely surprising, given that the flow velocity is non-zero over significant portions of the downstream sweep, even for $\lambda\:=\:3.4$ (figure \ref{flow time-average}). 

\section{Discussion}
\label{discussion}
\subsection{Generalization to Other Geometries}
To understand whether the observed performance trends in section \ref{results} generalize to a broader set of turbine geometries and flow conditions, we apply the upstream and downstream segment-averaging framework to 54 additional, unique, one-bladed turbine performance experiments from Hunt et al. \cite{hunt2023parametric,GeoData}. In these experiments, the authors independently considered 5 chord lengths ($c/r\:=\:0.25$ to $0.74$), 6 preset pitch angles ($\alpha_p\:=\:-2^\circ$ to $-12^\circ$), and 3 diameter-based Reynolds numbers ($Re_D\:=\:0.75\times10^5$ to $2.7\times10^5$). These experiments utilized the same blade aspect ratio, turbine diameter, and ``performance measurement'' test setup as described in section \ref{test rigs}, but the foil strut supports were replaced with circular plates to facilitate the different geometric variations. All performance data is processed as described in section \ref{test rigs}.

\begin{figure*}[t!]
    \centering
    \includegraphics[width=1\linewidth]{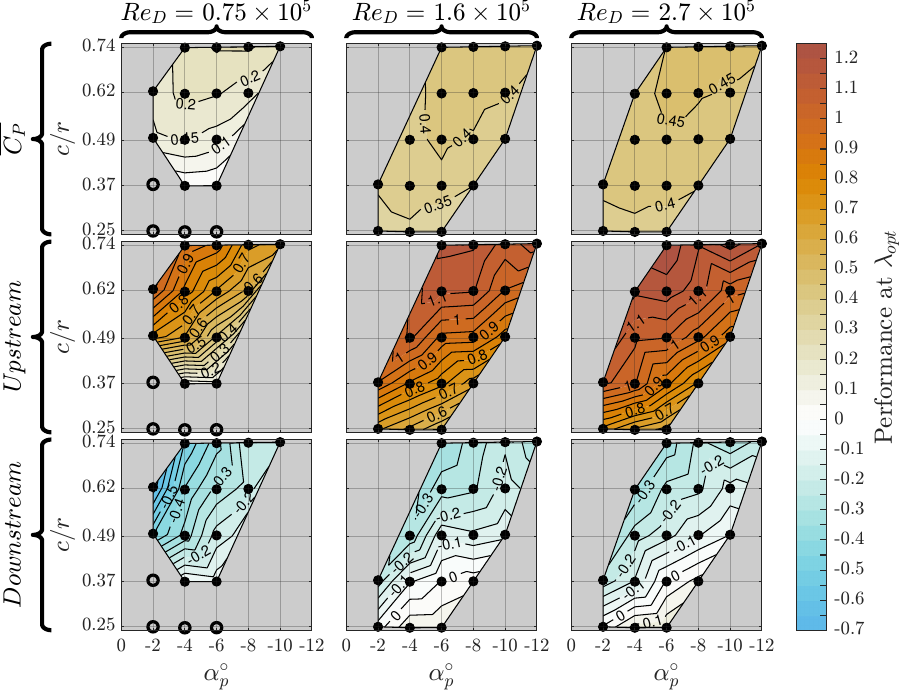}
    \caption{Total, upstream, and downstream segment-averaged performance at the optimal tip-speed ratio as a function of $c/r$, $\alpha_p$, and $Re_D$ for the 54 single-bladed combinations tested by Hunt et al. \cite{hunt2023parametric}. The optimal tip-speed ratio depends on the particular geometric configuration, but is largely invariant to $Re_D$. The unfilled circles indicate experiments that did not produce positive time-averaged performance for any tested $\lambda$.}
    \label{Additional Geo peaks}
\end{figure*}

\begin{figure*}[t!]
    \centering
    \includegraphics[width=1\linewidth]{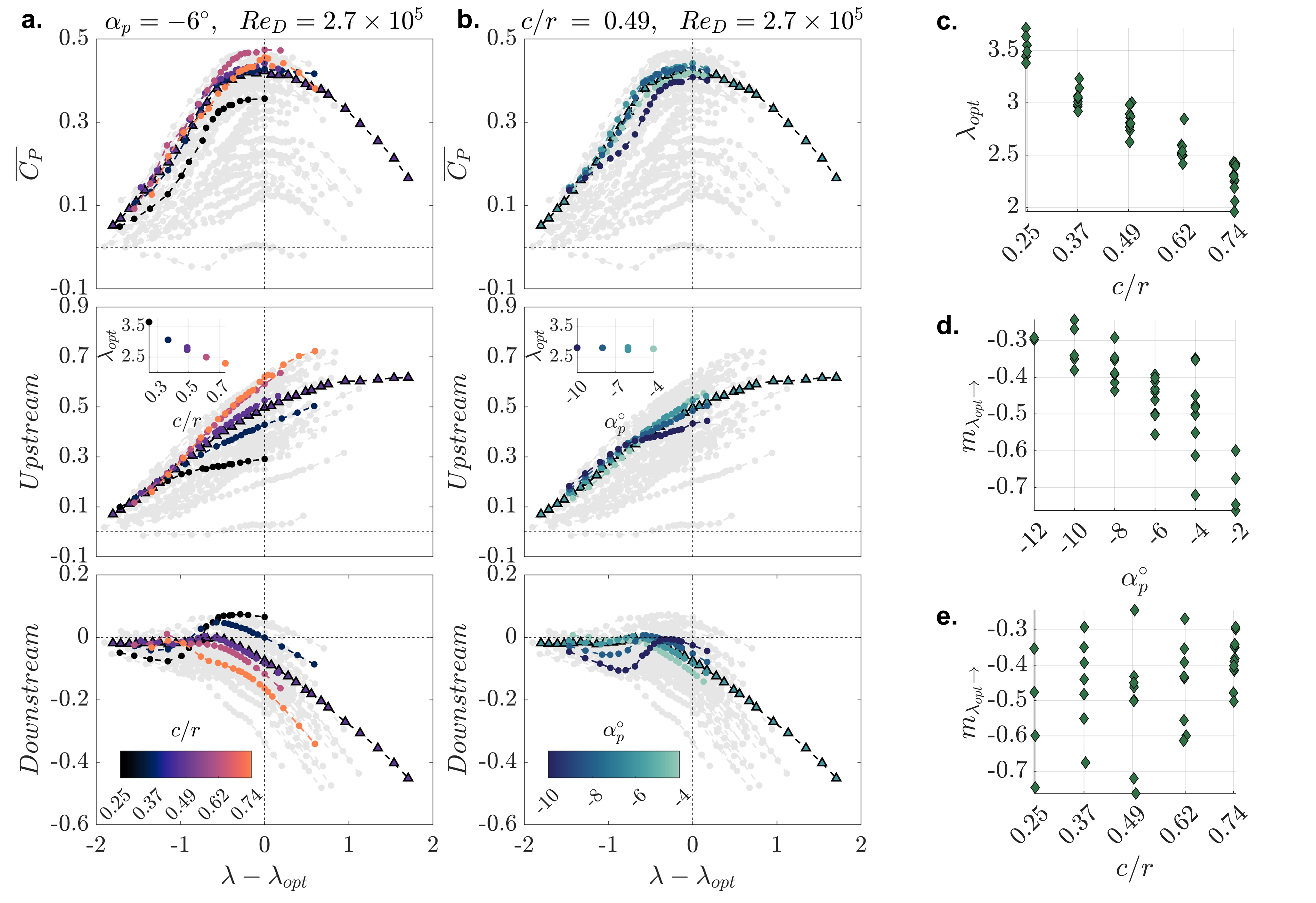}
    \caption{(a,b) Time and segment-averaged characteristic performance curves for all 55 unique turbine performance experiments. The triangles represent the one-bladed data collected for this work and the circles represent the 54 other one-bladed experiments from Hunt et al. \cite{hunt2023parametric}. The colored lines represent select data from two slices through the parameter space tested. In each column, a single parameter (a: $c/r$, and b: $\alpha_p$) is varied, as indicated by the color bar. The gray lines represent all of the other experiments. Experiments that never produce positive blade-level performance are omitted for clarity. The optimal tip-speed ratio, $\lambda_{opt}$, is used to align the performance peaks of all experiments and is plotted against the selected parameter in the insets. (c) Relationship between the chord-to-radius ratio and the optimal tip-speed ratio for one-bladed turbines. (d) Relationship between the preset pitch angle and the slope of the linear fit to the coefficient of performance per blade area past the performance peak. (e) Relationship between the chord-to-radius ratio and the slope of the linear fit of the coefficient of performance per blade area past the performance peak.}
    \label{Additional Geo}
\end{figure*}

Contours of segment-averaged performance within the upstream and downstream regions at the optimal tip-speed ratio for each $c/r\:-\:\alpha_p$ combination are shown in figure \ref{Additional Geo peaks}, with each column representing a different $Re_D$. Broadly, we observe contrasting performance trends between the upstream and downstream sweeps at the optimal tip-speed ratios. Optimal performance in the upstream sweep increases with $c/r$, but decreases as $\alpha_p$ becomes more negative, regardless of $c/r$. For both parameters, the opposite is true for the downstream sweep. As expected, optimal performance increases with $Re_D$ in both the upstream and downstream sweeps. The characteristic performance curves in figure \ref{Additional Geo}a,b show that the segment-averaged trends are consistent with the experiment detailed in section \ref{results} (triangle markers; $Re_D\:=\:2.3\times10^5$). Specifically, performance in the upstream sweep continues to increase well beyond the optimal tip-speed ratio for all experiments, such that the optimal tip-speed ratio is a consequence of the approximately linear degradation in downstream performance (figure \ref{Additional Geo}a,b). 

The optimal tip-speed ratio, as reported by Hunt et al. \cite{hunt2023parametric} and Rezaeiha et al. \cite{rezaeiha2018solidity}, is inversely proportional to turbine solidity (equation \ref{solidity}). This relationship is evident in figure \ref{Additional Geo}c, for which $N$ and $r$ are constant and $c$ varies. Rezaeiha et al. \cite{rezaeiha2018solidity} show that induction increases with solidity, so we hypothesize that larger $c/r$ turbines (higher solidity) have lower optimal tip-speed ratios because of increased induction that accelerates the convergence to the limiting case (figure \ref{lift drag}a). Beyond the optimal tip-speed ratio, the slope of downstream performance degradation is most closely correlated with the preset pitch angle (figure \ref{Additional Geo}d,e) and steepens for less negative preset pitch angles. This slope, $m_{\lambda_{opt}\rightarrow}$, is defined by the linear fit ($r^2>0.95$ for all cases) of $\overline{C_P} \frac{D}{c}$ (performance per blade area) for $\lambda\:\geq\:\lambda_{opt}$ when there are at least 3 points at and beyond $\lambda_{opt}$. The relationship between $m_{\lambda_{opt}\rightarrow}$ and $\alpha_p$ is unsurprising since the aerodynamic coefficients are implicit functions of the angle of attack which is influenced by the preset pitch angle. 

For two reasons, it is unlikely the limiting case (section \ref{Downstream Model}) is reached in the current dataset and/or our analytical model does not adequately describe all phases of the downstream sweep. First, for all geometries, the linear performance degradation beyond $\lambda_{opt}$ differs from the $\lambda^3$ dependency in the analytical model (equation \ref{CPlimit}). Second, in the framework of the limiting kinematics, $m_{\lambda_{opt}\rightarrow}$ should not steepen as $\alpha_p$ becomes less negative. For steady-state, high tip-speed ratio kinematics, if the performance were drag-dominated in the downstream sweep, a shallower slope would be expected for smaller preset pitch angles (i.e., less negative angle of attack, less drag, less negative $m_{\lambda_{opt}\rightarrow}$). The inverse trend suggests the combined influence of lift, drag, the pitching moment, and virtual camber affect this slope and requires further analysis.

The opposing performance trends between the upstream and downstream sweep with $c/r$ in figure \ref{Additional Geo}a are likely a consequence of upstream momentum extraction and/or different force regimes: lift vs. drag/moment dominated. In terms of momentum extraction, the largest chord-length blades (higher turbine solidity) perform best in the upstream sweep, meaning they may also remove more momentum from the flow, limiting momentum available to the downstream sweep. This is consistent with conclusions drawn by Rezaeiha et al. \cite{rezaeiha2018solidity}. 
In terms of different force regimes, the upstream sweep is lift-dominated for all tip-speed ratios (evidenced by positive performance), while the downstream sweep is drag/moment-dominated at high tip-speed ratios (evidenced by negative performance). Therefore, a larger blade chord (more blade area) is beneficial in the upstream sweep (increased lift force) but detrimental in the downstream sweep (increased drag force). Once again the pitching moment may either enhance or detract from detrimental torque in the downstream sweep. The significance of the pitching moment would depend on $c/r$ through virtual camber and incidence (affecting the pitching moment coefficient) and the moment magnitude (dependence on $Hc^2$). 

Opposing performance trends are also apparent between the upstream and downstream sweeps for varying $\alpha_p$ (figure \ref{Additional Geo}b). During the upstream sweep, for $\lambda\:-\:\lambda_{opt} < -1$, more negative preset pitch is slightly preferable, but at higher tip-speed ratios less negative preset pitch is preferable. The reverse is true for downstream performance, with less negative preset pitch angles preferable at low $\lambda$ and more negative preset pitch angles preferable at high $\lambda$. These opposing trends in the upstream and downstream regions are likely the result of the reversal of the pressure and suction sides of the blade as it transitions between each region. By our sign convention (equation \ref{nominal aoa}), this means negative preset pitch angles decrease the angle of attack magnitude in the upstream sweep, but increase it in the downstream sweep. As with $c/r$, these trends could also result from reductions in downstream momentum. For example, optimal preset pitch angles within the downstream sweep are generally detrimental to upstream power production, theoretically leaving more momentum available to the downstream portion of the rotation. The opposite would be true for optimal preset pitch angles within the upstream sweep.

All cases are at similar relatively low blockages, $10-11\%$, but higher blockages could affect the dynamics in the upstream and downstream sweeps. Blockage increases velocity through the rotor plane as a consequence of an increased pressure drop \cite{nishino2012efficiency}. Because of this, the observed upstream/downstream trends and the tip-speed ratio required to reach the limiting case may be functions of blockage. Experiments at much higher blockage ratios (e.g., $> 50\%$) \cite{BlockageData}, have the same qualitative trends observed here (e.g., power production in the upstream sweep continues well beyond the optimal tip-speed ratio while the downstream sweep begins to consume appreciable power).


\subsection{Improving Downstream Sweep Performance}
Performance in the upstream sweep continues to increase beyond the optimal tip-speed ratio (figure \ref{torque and perf}a). Therefore, even if it is not possible to produce power in the downstream sweep, it may be possible to improve overall turbine performance by reducing detrimental downstream torque. At high tip-speed ratios, at least two mechanisms could improve downstream torque:

(1) entrain flow or momentum into the rotor (reducing induction, increasing $\lambda_{lim}$) in either the upstream or downstream sweep.

(2) limit parasitic torque by reducing the drag coefficient, altering the pitching moment and/or increasing lift. We note that each of these methods will limit parasitic torque, but because of the directionality of lift at high tip-speed ratios, changes to drag or pitching moment may have a larger impact.

Control strategies such as active pitch control have the potential to improve downstream performance \cite{Mulleners2023activepitch}. Active pitch control strategies alter the turbine kinematics by varying the preset pitch angle throughout the rotation. This affects the amplitude, slope, and shape of the angle of attack trajectory and, therefore, the lift, drag and pitching moment coefficients. In the downstream sweep, the in-rotor velocity and the tangential velocity do not depend on the blade pitch angle, so active pitch control will not affect the relative velocity vector or the direction of lift and drag. However, the ability to augment the aerodynamic coefficients throughout the rotation is still important and each blade can be actuated independently, facilitating individualized control between the upstream and downstream sweeps. 

\section{Conclusion}
To date, research on optimizing cross-flow turbine performance has focused on the hydrodynamics of the upstream blade sweep, where most of the power is produced. In this work, we utilize turbine performance and torque measurements in concert with phase-locked PIV flow fields to directly characterize the dynamics and performance of the downstream sweep and to investigate the influence of the downstream sweep on overall cross-flow turbine performance. 
Performance data are compared for 55 unique combinations of chord-to-radius ratio ($c/r\:=\:0.25$ to $0.74$), preset pitch angle ($\alpha_p\:=\:-2^\circ$ to $-12^\circ$), and Reynolds number ($Re_D\:=\:0.75\times10^5$ to $2.7\times10^5$). Corresponding flow fields are presented for one of these conditions ($c/r\:=\:0.47$, $\alpha_p\:=\:-6^\circ$, and $Re_D\:=\:2.2\times10^5$) at tip-speed ratios corresponding to near-optimal, $\lambda\:=\:2.4$, and sub-optimal, $\lambda\:=\:1.4, 3.4$, turbine performance. 

Across all experiments, the time-averaged performance in the upstream sweep continues to increase well beyond the optimal tip-speed ratio for complete cycles. In contrast, beyond the optimal tip-speed ratio, time-averaged performance in the downstream sweep decreases approximately linearly, at a faster rate than the upstream performance increases. This indicates that the optimal tip-speed ratio and maximum turbine efficiency are strongly influenced by performance degradation in the downstream sweep. During the upstream sweep, the magnitude of the torque peak converges at higher tip-speed ratios, representing a balance between the lift, drag, and pitching moment coefficients, the relative velocity, and the projection of the aerodynamic forces into the direction of rotation. Therefore, the increased performance at high tip-speed ratios in the upstream sweep is the result of the tip-speed ratio multiplier and not higher torque production. In contrast, as tip-speed ratio increases, parasitic torque is continuously more detrimental in the downstream sweep. Additionally, we identify contrasting performance trends between the upstream and downstream sweep with $c/r$ and $\alpha_p$, indicating that geometric parameters can have different impacts between these regions. Specifically, we observe that it is common that the parameters that are optimal for upstream performance are not optimal for overall turbine performance. 

Flow field measurements suggest faster flow re-attachment during the downstream sweep at higher tip-speed ratios. While this would be expected to reduce drag, we observe increasingly parasitic torque. These trends indicate the near-blade dynamics are not likely the mechanism most responsible for the monotonic downstream performance degradation. Instead, reductions to available momentum and changes to the blade kinematics due to induction are most explanatory. In addition to reducing the momentum available to the downstream sweep for power production, induction limits how effectively blades convert the available momentum into torque. Specifically, induction drives the convergence to a limiting case where the relative velocity incident on the blade is equal to the tangential velocity, such that lift acts normal to the direction of rotation and drag directly opposes rotation. While it is unlikely the limiting case is reached for any of the experiments investigated in this work, lift and drag act in increasingly unfavorable directions as induction and the tip-speed ratio increase, diminishing torque production from lift and increased parasitic drag. A pitching moment arising from rotation-induced virtual camber may further oppose rotation, but more detailed analysis is required to understand its relative magnitude and parametric dependencies. 

Improvements to downstream torque will increase overall turbine torque generation and the optimal tip-speed ratio -- if this can be done without degrading upstream torque. Because the performance coefficient is the product of the torque coefficient and the tip-speed ratio, any improvements in torque are amplified in performance by the multiplication with higher optimal tip-speed ratios. Overall, the downstream sweep is critical to cross-flow turbine performance, and understanding it is necessary to improve power production, as well as to decipher the potential benefits of alternative turbine geometries and advanced control strategies. 


%
%

%

\section{Supplemental Material}
A full derivation of the limiting case coefficient of performance (equation \ref{CPlimit}) is available in the supplemental material.

\begin{acknowledgments}
The authors thank the Alice C. Tyler Charitable Trust for supporting the research facility and acknowledge the substantial contributions by Benjamin Strom, Hannah Ross, Carl Stringer, Erik Skeel, and Craig Hill to the development and upgrades of the experimental setup and code base.
\subsection*{Funding}
Financial support was received from the United States Department of Defense Naval Facilities Engineering Systems Command and through the National Science Foundation Graduate Research Fellowship Program. 
\subsection*{Competing interests}
The authors have no relevant financial or non-financial interests to disclose.
\subsection*{Availability of data and materials}
Data and processing codes are available upon request.
\end{acknowledgments}

\bibliography{PAPER}

\end{document}